 \documentclass[preprint]{aastex}

\newcommand{\hrcs}{\mbox{HRC-S}}
\shorttitle{First light with LETGS}
\shortauthors{Brinkman et al.}

\begin{document}
\title{First Light Measurements of Capella with the Low Energy Transmission
	Grating Spectrometer aboard the Chandra X-ray Observatory}

\author{A.C.~Brinkman, 
	C.J.T.~Gunsing,
	J.S.~Kaastra,
	R.L.J.~van~der~Meer,
	R.~Mewe,
	F.~Paerels \altaffilmark{1},
	A.J.J.~Raassen \altaffilmark{2},
	J.J.~van~Rooijen}
\affil{Space Research Organization of the Netherlands (SRON), 
	Sorbonnelaan~2, 3584~CA\ \ Utrecht, The~Netherlands}

\author{H.~Br\"{a}uninger, 
	W.~Burkert,
	V.~Burwitz,
	G.~Hartner,
	P.~Predehl}
\affil{Max-Planck-Institut f\"{u}r Extraterrestrische Physik (MPE), 
	Postfach~1603, D-85740~Garching, Germany}

\author{J.-U.~Ness, J.H.M.M.~Schmitt}
\affil{Universit\"at Hamburg, Gojenbergsweg~122, D-21029 Hamburg, Germany}

\author{J.J.~Drake,
	O.~Johnson,
	M.~Juda,
	V.~Kashyap,
	S.S.~Murray,
	D.~Pease,
	P.~Ratzlaff,
	B.J.~Wargelin}
\affil{Harvard-Smithsonian Center for Astrophysics, 60~Garden Street, 
	Cambridge, MA~02138, USA}

\altaffiltext{1}{Present address, Columbia University, New York, NY, USA}
\altaffiltext{2}{Also at Astronomical Institute "Anton Pannekoek", 
Kruislaan~403, 1098~SJ\ \ Amsterdam, The Netherlands}

\begin{abstract}
We present the first X-ray spectrum obtained by the Low Energy
Transmission Grating Spectrometer (LETGS) aboard the Chandra X-ray 
Observatory.  The spectrum is of Capella and covers a wavelength
range of 5--175~\AA\ (2.5--0.07~keV).  The measured wavelength
resolution, which is in good agreement with ground calibration, is
$\Delta \lambda \simeq$ 0.06~\AA\ (FWHM).
Although in-flight calibration of the LETGS is in progress, the 
high spectral resolution and unique wavelength coverage of the LETGS
are well demonstrated by the results from Capella, a coronal 
source rich in spectral emission lines.
While the primary purpose of this letter is to demonstrate the spectroscopic
potential of the LETGS, we also briefly present some preliminary astrophysical
results.  We discuss plasma parameters derived from line ratios in
narrow spectral bands, such as the electron density diagnostics of the
He-like triplets of carbon, nitrogen, and oxygen, as well as resonance
scattering of the strong Fe~XVII line at 15.014~\AA.

\end{abstract}


\section{Introduction}

The LETGS consists of three components of the Chandra Observatory: 
the High Resolution Mirror Assembly (HRMA) \citep{Spe97}, 
the Low Energy Transmission Grating (LETG) \citep{Brink87,Brink97,Pre97}, 
and the spectroscopic array of the High Resolution Camera (\hrcs) 
\citep{Mur97}. 
The LETG, designed and manufactured in a collaborative effort
of SRON in the Netherlands and MPE in Germany, 
consists of a toroidally shaped structure which supports
180~grating modules. Each module holds three 1.5-cm~diameter
grating facets which have a line density of 1008~lines/mm. 
The three flat detector elements of the \hrcs, each 10~cm long and 2~cm 
wide, are tilted to approximate the Rowland focal surface
at all wavelengths, assuring a nearly coma-free spectral image. 
The detector can be moved in the cross-dispersion direction and along the 
optical axis, to optimize the focus for spectroscopy. 
\footnote{ 
  Further information on LETGS components is found in the 
  AXAF Observatory Guide (\url{http://asc.harvard.edu/udocs/})
  and at the Chandra X-ray Center calibration webpages 
  (\url{http://asc.harvard.edu/cal/}).
}
 
An image of the LETG spectrum is focused on
the \hrcs\ with zeroth order at the focus position and 
dispersed positive and negative orders symmetric on either side of it. 
The dispersion is 1.15~\AA/mm in first spectral order.
The spectral width in the cross-dispersion direction is minimal
at zeroth order and increases at larger wavelengths due to the
intrinsic astigmatism of the Rowland circle spectrograph. The extraction of
the spectrum from the image is done by applying a spatial filter around the 
spectral image and constructing a histogram of counts vs. position along the
dispersion direction.
The background is estimated from areas on the detector away
from the spectral image 
and can be reduced by filtering events by pulse-height.
   
\section{First Light Spectrum}

Capella is a binary system at a distance of 12.9~pc consisting of G8 and G1
giants with an orbital period of 104~days \citep{Hum94}.  It is
the brightest quiescent coronal X-ray source in the sky after the Sun,
and is therefore an obvious line source candidate for first light and for 
instrument calibration.
X rays from Capella were discovered in 1975 \citep{Cat75,Mew75} 
and subsequent satellite observations provided evidence for a 
multi-temperature component plasma (e.g. \citet{Mew91a} for references).
Recent spectra were obtained with EUVE longward of 70~\AA\ 
with a resolution of about 0.5~\AA\ \citep{Dup93,Sch95}.

\begin{figure}
  \epsscale{0.8}
  \plotone{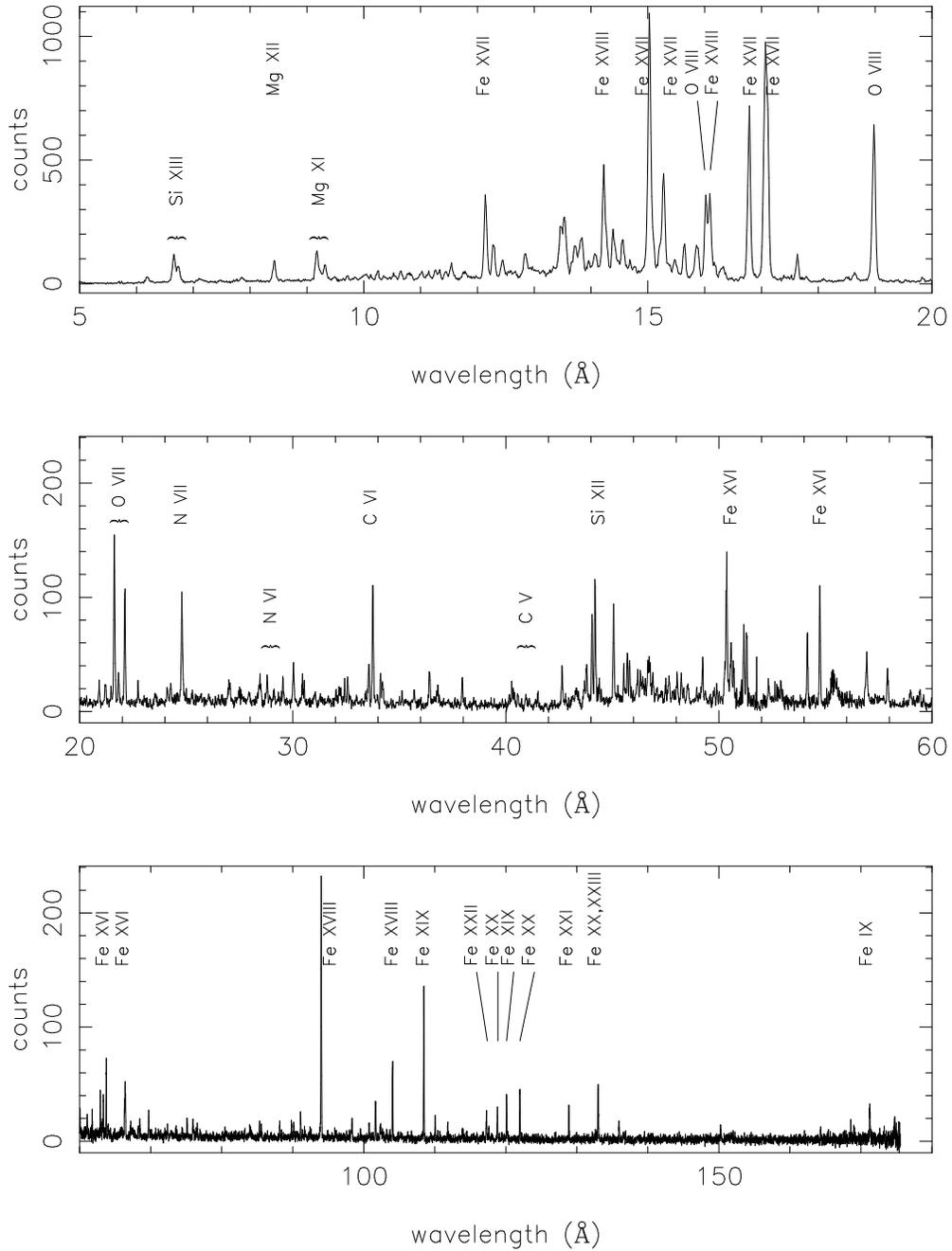}
  \caption
  {The complete LETGS spectrum of Capella, split into three parts for clarity. 
   Note the difference in x and y scale for the three parts. 
   Indicated in the plot are the triplets discussed in the text
   and a selection of the Fe lines at longer wavelengths.
   The hundred strongest lines are listed in Table~\ref{tab1}.
   \label{spec_img}
  }
\end{figure}

The LETG First Light observation of Capella was performed on 6~September
1999 (00h27m UT -- 10h04m UT) with LETG and \hrcs. 
For the analysis we use a composite of six observations obtained 
in the week after first light, with a total observing time of 95~ksec.
The \hrcs\ output was processed through standard pipeline processing.
For LETG/\hrcs\ events, only the product of the
wavelength and diffraction order is known because no diffraction order
information can be extracted.  Preliminary analysis of the pipeline
output immediately revealed a beautiful line-rich spectrum.  The
complete background-subtracted, negative-order spectrum between 5~and
175~\AA\ is shown in Fig.~\ref{spec_img}.
  \notetoeditor{Figure~\ref{spec_img} (PostScript file spec\_img.ps) and
  Figure~\ref{OVII} (PostScript file OVII.ps) can be printed in the text.}
Line identifications were made using previously measured
and/or theoretical wavelengths from the literature.  
The most prominent lines are listed in Table~\ref{tab1}.

The spectral resolution $\Delta \lambda$ of the LETGS is nearly
constant when expressed in wavelength units, and therefore the
resolving power $\lambda / \Delta \lambda$ is greatest at long
wavelengths.  With the current uncertainty of the LETGS wavelength
scale of about 0.015~\AA, this means that the prominent lines at 150~and 
171~\AA\ could be used to measure Doppler shifts as small as 30~km/sec,
such as may occur during stellar-flare mass ejections, once
the absolute wavelength calibration of the instrument has been
established.  This requires, however, that line rest-frame wavelengths
are accurately known and that effects such as the orbital velocity of
the Earth around the Sun are taken into account.  Higher-order lines,
such as the strong O VIII Ly$\alpha$ line at 18.97~\AA, which is seen out to
6th order, can also be used.

\section{Diagnostics}

A quantitative analysis of the entire spectrum by multi-temperature fitting or 
differential emission measure modeling yields a detailed thermal 
structure of the corona, but this requires accurate detector efficiency 
calibration which has not yet been completed.
However, some diagnostics based on intensity ratios of lines lying closely 
together can already be applied. In this letter we consider the helium-like 
line diagnostic and briefly discuss the resonance scattering in the 
Fe~XVII 15.014~\AA\ line.

\subsection{Electron Density \& Temperature Diagnostics} 

Electron densities, $n_e$, can be measured using density-sensitive 
spectral lines originating from metastable levels, 
such as the forbidden~($f$) $2^3S\to 1^1S$ line in helium-like ions.
This line and the associated resonance~($r$) $2^1P\to 1^1S$ and
intercombination~($i$) $2^3P\to 1^1S$ line make up the so-called 
helium-like "triplet" lines \citep{Gab69,Pra82,Mew85}.
The intensity ratio $(i+f)/r$ varies with electron
temperature, T, but more importantly, the ratio $i/f$ varies
with $n_e$ due to the collisional coupling between the
$2^3S$ and $2^3P$ level.

\begin{figure}
  \epsscale{0.8}
  \plotone{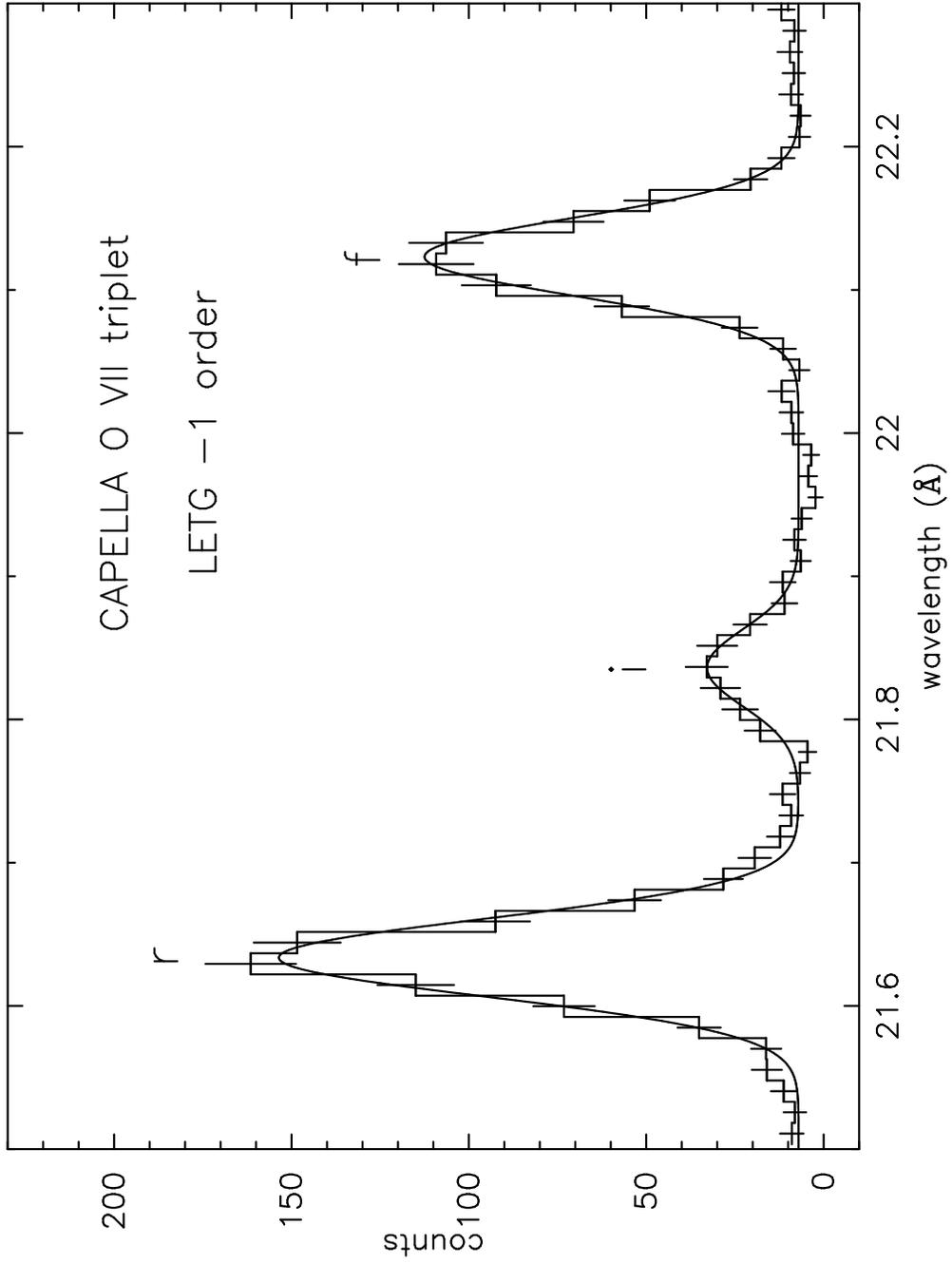}
  \caption
  {The Oxygen VII triplet in the LETGS --1~order spectrum with the 
   resonance~(r), the forbidden~(f), and the intercombination~(i) line.
   The measured ratios of these lines (from the fitted curve) are given in 
   the text.
   \label{OVII}
  }
\end{figure}

The LETGS wavelength band contains the He-like triplets from C, N, O, Ne, Mg, 
and Si ($\sim$ 40, 29, 22, 13.5, 9.2, and 6.6~\AA, respectively). 
However, the Si and Mg triplets are not sufficiently resolved and the 
Ne~IX triplet is too heavily blended with iron and nickel lines 
for unambiguous density analysis.
The O~VII lines are clean (see Fig.~\ref{OVII}) and the C~V and N~VI lines 
can be separated from the blends by simultaneous fitting of all lines.
These triplets are suited to diagnose plasmas in 
the range $n_e$ = 10$^8$--10$^{11}$~cm$^{-3}$ 
and $T$~$\sim$~1--3~MK. 
For the C, N, and O triplets the measured $i/f$ ratios are
$0.38\pm 0.14$, 
$0.52\pm 0.15$, and
$0.250\pm 0.035$, respectively, 
which imply \citep{Pra82} $n_e$ (in $10^9$~cm$^{-3}$) = 
$2.8\pm 1.3$, 
$6\pm 3$, and 
$\la$~5 (1$\sigma$ upper limit), respectively, for typical temperatures as 
indicated by the $(i+f)/r$ ratios of 1, 1, and 3~MK, respectively. 
This concerns the lower temperature part of a multi-temperature structure 
which also contains a hot ($\sim$6--8~MK), 
and dense ($\ga$~10$^{12}$~cm$^{-3}$) compact plasma component 
(see Section~\ref{resonance}).
The derived densities are comparable to those of active regions on the 
Sun with a temperature of a few~MK. 
Fig.~\ref{OVII} shows a fit to the O~VII triplet measured in the --1~order. 
The He-like triplet diagnostic, which was first applied to the Sun
(e.g., \citet{Act72,Wol83}) has now for the first time been applied 
to a star other than the Sun.

The long-wavelength region of the LETGS between 90~and 150~\AA\ contains 
a number of density-sensitive lines from $2\ell$--$2\ell'$ transitions in 
the Fe-L ions Fe~XX--XXII which provide density diagnostics 
for relatively hot ($\ga$~5~MK) and dense ($\ga$~10$^{12}$~cm$^{-3}$) plasmas 
\citep{Mew85,Mew91b,Brick95}. These have been applied in a few cases to 
EUVE spectra of late-type stars and in the case of Capella have 
suggested densities more than two orders of magnitude higher than 
found here for cooler plasma \citep{Dup93,Sch95}. 
These diagnostics will also be applied to the LETGS spectrum as soon 
as the long-wavelength efficiency calibration is established.

\subsection{The 15--17~\AA\ region: resonance scattering of Fe~XVII?}
\label{resonance}

Transitions in Ne-like Fe~XVII yield the strongest emission lines 
in the range 15--17~\AA\ (cf. Fig.~\ref{spec_img}).  
In principle, the optical depth, $\tau$, in the 15.014~\AA\ line
can be obtained by 
applying a simplified escape-factor model to the ratio of the 
Fe~XVII~15.014~\AA\ resonance line with a large oscillator strength 
to a presumably optically thin Fe~XVII line with a small oscillator strength. 
We use the 15.265~\AA\ line because the 16.780~\AA\ line 
can be affected by radiative cascades \citep{Lie99}.
Solar physicists have used this technique to derive the density in active 
regions on the Sun (e.g., \citet{Sab99,Phi96,Phi97}). 

Various theoretical models predict 15.014/15.265~ratio values 
in the range 3.3--4.7 with only a slow variation ($\la$~5\%) 
with temperature or energy in the region 2--5~MK or 0.1--0.3~keV 
\citep{Bro98,Bha92}. The fact that most ratios observed in the Sun 
typically range from 1.5--2.8 (\citet{Bro98}, and references above), 
significantly lower than the theoretical ratios, supports 
claims that in solar active regions the 15.014~\AA\ line is affected by 
resonant scattering. 
The 15.014/15.265~ratio which was recently measured in the Livermore 
Electron Beam Ion Trap (EBIT) \citep{Bro98} and ranges from 2.77--3.15 
(with individual uncertainties of about $\pm~0.2$) at energies between 
0.85--1.3~keV, is significantly lower than calculated values. Although the 
EBIT results do not include probably minor contributions from processes such 
as dielectronic recombination satellites and resonant excitation, 
this may imply that the 
amount of solar scattering has been overestimated in past analyses. 
Our measured ratio Fe~XVIII 16.078~\AA/Fe~XVII 15.265~\AA\ 
gives a temperature of $\sim$6~MK and the 
photon flux ratio 15.014/15.265 is measured to be 2.64$\pm 0.10$. 
If we compare this to the recent EBIT results we conclude that there is 
little or no evidence for opacity effects in the 15.014~\AA\ line seen 
in our Capella spectrum.

\section{Conclusion}

The Capella measurements with LETGS show a rich spectrum with
excellent spectral resolution ($\Delta\lambda \simeq $0.06~\AA, FWHM).
About 150~lines have been identified of which the brightest hundred
are presented in Table~\ref{tab1}. The high-resolution spectra of the Chandra
grating spectrometers allow us to carry out direct density
diagnostics, using the He-like triplets of the most abundant elements
in the LETGS-band, which were previously only possible for the Sun.  
Density estimates based on C, N and O He-like complexes indicate 
densities typical of solar active regions and some two or more 
orders of magnitude lower than density estimates for the hotter 
($>$5~MK) plasma obtained from EUVE spectra.  
A preliminary investigation into the effect of resonance scattering in
the Fe~XVII line at 15.014~\AA\ showed no clear evidence for opacity
effects.  After further LETGS in-flight calibration it is expected
that relative Doppler velocities of the order of 30~km/s will be
detectable at the longest wavelengths.

\acknowledgments

The LETGS data as presented here could only be produced after
dedicated efforts of many people for many years. Our special gratitude
goes to the technical and scientific colleagues at SRON, MPE and their
subcontractors for making such a superb LETG and to the colleagues at
many institutes for building the payload.  Special thanks goes to the
many teams who made Chandra a success, particularly the project
scientist team, headed by Dr. Weisskopf, the MSFC project team, headed
by Mr. Wojtalik, the TRW industrial teams and their subcontractors, 
the Chandra observatory team, headed by Dr. Tananbaum, 
and the crew of Space Shuttle flight STS-93.

JJD, OJ, MJ, VK, SSM, DP, PR, and BJW were supported by Chandra X-ray 
Center NASA contract NAS8-39073 during the course of this research.

\clearpage

\clearpage

\begin{deluxetable}{r@{}l@{}r@{ }c@{ }rl@{ }ll}
\tabletypesize{\footnotesize}
\tablecaption{
  Comparison of measured and theoretical values of the strongest lines in the
  Capella spectrum as shown in Fig.~\ref{spec_img}.
  \label{tab1}}
\tablewidth{0pt}
\tablehead{
  \multicolumn{2}{l}{$\lambda_{\rm obs}$} & 
  \colhead{$\lambda_{\rm pred}$} &
  \colhead{$\log(T_{m})$} & 
  \colhead{I} & 
  \multicolumn{2}{l}{Ion} & 
  \colhead{line ID}
}
\startdata
  6.65 &   &   6.65 &  7.00 &  5.1 &  Si &   XIII & He4w \\
  6.74 &   &   6.74 &  7.00 &  2.9 &  Si &   XIII & He6z \\
  8.42 &   &   8.42 &  7.00 &  4.6 &  Mg &    XII & H1AB \\
  9.16 &   &   9.17 &  6.80 &  6.2 &  Mg &     XI & He4w \\
  9.31 &   &   9.32 &  6.80 &  3.1 &  Mg &     XI & He6z \\
 11.54 & b &  11.55 &  6.60 &  3.5 &  Ne &     IX & He3A \\
   ... &   &  11.53 &  6.85 &      &  Fe &  XVIII & F22 \\
 12.14 & b &  12.13 &  6.80 & 16.8 &  Ne &      X & H1AB \\
   ... &   &  12.12 &  6.75 &      &  Fe &   XVII & 4C \\
 12.27 & b &  12.26 &  6.75 &  6.6 &  Fe &   XVII & 4D \\
   ... &   &  12.29 &  7.00 &      &  Fe &    XXI & C13 \\
 12.43 &   &  12.43 &  6.70 &  3.5 &  Ni &    XIX & Ne5 \\
 12.84 & b &  12.83 &  7.00 &  4.9 &  Fe &     XX & N16 \\
   ... &   &  12.85 &  7.00 &      &  Fe &     XX & N15 \\
 13.46 &   &  13.45 &  6.60 &  9.7 &  Ne &     IX & He4w \\
 13.53 & b &  13.52 &  6.90 & 11.5 &  Fe &    XIX & O1-68 \\
   ... &   &  13.51 &  6.90 &      &  Fe &    XIX & O1-71 \\
   ... &   &  13.55 &  6.90 &      &  Ne &     IX & He5xy \\
 13.71 &   &  13.70 &  6.60 &  6.6 &  Ne &     IX & He6z \\
 13.82 & b &  13.83 &  6.70 &  7.5 &  Fe &   XVII & 3A \\
   ... &   &  13.84 &  6.90 &      &  Fe &    XIX & O1-50 \\
 14.07 &   &  14.06 &  6.70 &  4.2 &  Ni &    XIX & Ne8AB \\
 14.22 &   &  14.21 &  6.80 & 18.0 &  Fe &  XVIII & F1-56,55 \\
 14.27 &   &  14.26 &  6.80 &  5.3 &  Fe &  XVIII & F1-52,53 \\
 14.38 & b &  14.38 &  6.80 &  6.2 &  Fe &  XVIII & F12 \\
   ... &   &  14.36 &  6.80 &      &  Fe &  XVIII & F2-57,58 \\
 14.56 & b &  14.54 &  6.80 &  5.3 &  Fe &  XVIII & F10 \\
   ... &   &  14.56 &  6.80 &      &  Fe &  XVIII & F9 \\
 15.02 &   &  15.01 &  6.70 & 44.2 &  Fe &   XVII & 3C \\
 15.18 & b &  15.18 &  6.60 &  3.4 &   O &   VIII & H3 \\
   ... &   &  15.21 &  6.90 &      &  Fe &    XIX & O4 \\
 15.27 &   &  15.27 &  6.70 & 16.7 &  Fe &   XVII & 3D \\
 15.46 &   &  15.46 &  6.70 &  3.1 &  Fe &   XVII & 3E \\
 15.64 &   &  15.63 &  6.80 &  6.2 &  Fe &  XVIII & F7 \\
 15.83 &   &  15.83 &  6.80 &  4.3 &  Fe &  XVIII & F6 \\
 15.88 &   &  15.87 &  6.80 &  4.4 &  Fe &  XVIII & F5 \\
 16.02 & b &  16.01 &  6.60 & 14.6 &   O &   VIII & H2 \\
   ... &   &  16.00 &  6.80 &      &  Fe &  XVIII & F4 \\
 16.08 & b &  16.08 &  6.80 & 16.0 &  Fe &  XVIII & F3 \\
   ... &   &  16.11 &  6.90 &      &  Fe &    XIX & O2 \\
 16.30 & b &  16.34 &  6.70 &  2.2 &  Fe &   XVII & E2L \\
   ... &   &  16.31 &  6.80 &      &  Fe &  XVIII & F3-62 \\
 16.78 &   &  16.78 &  6.70 & 27.9 &  Fe &   XVII & 3F \\
 17.05 &   &  17.06 &  6.70 & 30.5 &  Fe &   XVII & 3G \\
 17.10 &   &  17.10 &  6.70 & 29.5 &  Fe &   XVII & M2 \\
 17.62 &   &  17.63 &  6.80 &  4.4 &  Fe &  XVIII & F1 \\
 18.62 & b &  18.63 &  6.80 &  2.0 &  Mg &     XI & He6z(2) \\
   ... &   &  18.63 &  6.30 &      &   O &    VII & He3A \\
 18.96 &   &  18.97 &  6.50 & 28.7 &   O &   VIII & H1AB \\
 21.61 &   &  21.60 &  6.30 &  6.5 &   O &    VII & He4w(r) \\
 21.82 &   &  21.80 &  6.30 &  1.1 &   O &    VII & He5xy(i) \\
 22.11 &   &  22.10 &  6.30 &  4.5 &   O &    VII & He6z(f) \\
 24.79 &   &  24.78 &  6.30 &  4.4 &   N &    VII & H1AB \\
 28.78 &   &  28.79 &  6.20 &  1.1 &   N &     VI & He4w \\
 29.52 &   &  29.53 &  6.20 &  0.9 &   N &     VI & He6z \\
 30.02 &   &  30.03 &  6.70 &  1.8 &  Fe &   XVII & 3C(2) \\
 33.74 &   &  33.74 &  6.10 &  4.9 &   C &     VI & H1AB \\
 34.10 &   &  34.10 &  6.70 &  1.5 &  Fe &   XVII & 3G(2) \\
 34.20 &   &  34.20 &  6.70 &  1.1 &  Fe &   XVII & M2(2) \\
 36.40 & b &  36.37 &  6.70 &  1.4 &  Fe &   XVII & 4C(3) \\
   ... &   &  36.40 &  6.30 &      &   S &    XII & B6A \\
 37.94 &   &  37.95 &  6.50 &  1.1 &   O &   VIII & H1AB(2) \\
 44.03 & b &  44.02 &  6.30 &  3.3 &  Si &    XII & Li6A \\
   ... &   &  44.05 &  6.10 &      &  Mg &      X & Li2 \\
 44.16 &   &  44.17 &  6.30 &  4.9 &  Si &    XII & Li6B \\
 45.03 &   &  45.04 &  6.70 &  4.2 &  Fe &   XVII & 3C(3) \\
 45.68 &   &  45.68 &  6.30 &  1.9 &  Si &    XII & Li7A \\
 50.31 &   &  50.35 &  6.50 &  5.3 &  Fe &    XVI & Na6A \\
 50.55 & b &  50.53 &  6.20 &  2.2 &  Si &      X & B6A \\
   ... &   &  50.56 &  6.50 &      &  Fe &    XVI & Na6B \\
 51.15 &   &  51.17 &  6.70 &  2.7 &  Fe &   XVII & 3G(3) \\
 51.27 &   &  51.30 &  6.70 &  2.9 &  Fe &   XVII & M2(3) \\
 54.12 &   &  54.14 &  6.50 &  2.9 &  Fe &    XVI & Na7B \\
 54.71 &   &  54.73 &  6.50 &  4.4 &  Fe &    XVI & Na7A \\
 56.89 &   &  56.92 &  6.50 &  1.8 &   O &   VIII & H1AB(3) \\
 60.04 &   &  60.06 &  6.70 &  1.3 &  Fe &   XVII & 3C(4) \\
 62.84 &   &  62.88 &  6.50 &  2.0 &  Fe &    XVI & Na8B \\
 63.68 &   &  63.72 &  6.50 &  2.9 &  Fe &    XVI & Na8A \\
 66.37 &   &  66.37 &  6.50 &  2.7 &  Fe &    XVI & Na9A \\
 68.20 &   &  68.22 &  6.70 &  1.0 &  Fe &   XVII & 3G(4) \\
 68.40 &   &  68.40 &  6.70 &  1.2 &  Fe &   XVII & M2(4) \\
 75.06 &   &  75.07 &  6.70 &  0.8 &  Fe &   XVII & 3C(5) \\
 75.87 &   &  75.89 &  6.50 &  0.9 &   O &   VIII & H1AB(4) \\
 85.24 &   &  85.28 &  6.70 &  0.8 &  Fe &   XVII & 3G(5) \\
 85.44 &   &  85.50 &  6.70 &  0.6 &  Fe &   XVII & M2(5) \\
 90.08 &   &  90.08 &  6.70 &  1.0 &  Fe &   XVII & 3C(6) \\
 93.91 &   &  93.92 &  6.80 & 12.4 &  Fe &  XVIII & F4A \\
 94.84 &   &  94.87 &  6.50 &  0.4 &   O &   VIII & H1AB(5) \\
101.55 &   & 101.55 &  6.90 &  2.5 &  Fe &    XIX & O6B \\
102.30 &   & 102.33 &  6.70 &  0.8 &  Fe &   XVII & 3G(6) \\
102.57 &   & 102.60 &  6.70 &  0.4 &  Fe &   XVII & M2(6) \\
103.94 &   & 103.94 &  6.70 &  4.4 &  Fe &  XVIII & F4B \\
108.35 &   & 108.37 &  6.90 &  6.1 &  Fe &    XIX & O6A \\
113.79 &   & 113.84 &  6.50 &  0.5 &   O &   VIII & H1AB(6) \\
117.14 &   & 117.17 &  7.10 &  1.2 &  Fe &   XXII & B11 \\
118.69 &   & 118.66 &  7.00 &  1.4 &  Fe &     XX & N6C \\
119.99 &   & 120.00 &  6.90 &  1.8 &  Fe &    XIX & O6D \\
121.86 &   & 121.83 &  7.00 &  2.0 &  Fe &     XX & N6B \\
128.74 &   & 128.74 &  7.00 &  1.6 &  Fe &    XXI & C6A \\
132.86 & b & 132.85 &  7.00 &  4.0 &  Fe &     XX & N6A \\
   ... &   & 132.85 &  7.10 &      &  Fe &  XXIII & Be13A \\
150.09 &   & 150.10 &  5.50 &  0.5 &   O &     VI & Li5AB \\
171.06 &   & 171.08 &  5.80 &  2.2 &  Fe &     IX & A4 \\
\enddata

\tablecomments{
$\lambda_{obs},\lambda_{pred}$: observed and predicted line wavelengths (\AA);
$T_{m}$ = temperature (in K) of maximum line formation; 
I = raw line intensity in $10^{-3}$~counts/sec.  Note that these
are included to illustrate approximate observed relative line strengths
and do not represent definitive measurements. \\
b = blend with indication of a the most prominent
    lines in order of estimated decreasing strength. \\
ID = line identification; number in parentheses (m)
     indicates diffraction order $m>1$. \\ 
For wavelengths and line identifications see 
\citet{Phi99} (solar and laboratory measurements between 5--20~\AA),
\citet{Mas84} (solar observations between 90--175~\AA), and 
\citet{Mew85,Mew95} (complete wavelength range in MEKAL).
}

\end{deluxetable}

\end{document}